\documentclass[aip,jcp,amsmath,amssymb,preprint]{revtex4-1}
\usepackage{tabularx,graphicx,subfigure}
\usepackage{dcolumn}
\usepackage{bm}
\usepackage{color}

\begin{document}

\title{First-Principles Equation of State and Electronic Properties of Warm Dense Oxygen}

\author{K. P. Driver}
 \affiliation{Department of Earth and Planetary Science, University of California, Berkeley, California 94720, USA}
 \email{kdriver@berkeley.edu}
 \homepage{http://militzer.berkeley.edu/~driver/}

\author{F. Soubiran}
 \affiliation{Department of Earth and Planetary Science, University of California, Berkeley, California 94720, USA}

\author{Shuai Zhang}
 \affiliation{Department of Earth and Planetary Science, University of California, Berkeley, California 94720, USA}

\author{B. Militzer}
 \affiliation{Department of Earth and Planetary Science, University of California, Berkeley, California 94720, USA}
 \affiliation{Department of Astronomy, University of California, Berkeley, California 94720, USA}

\date{\today}

\begin{abstract}
  We perform all-electron path integral Monte Carlo (PIMC) and density
  functional theory molecular dynamics (DFT-MD) calculations to
  explore warm dense matter states of oxygen. Our simulations cover a
  wide density-temperature range of $1-100$~g$\,$cm$^{-3}$ and
  $10^4-10^9$~K.  By combining results from PIMC and DFT-MD, we are
  able to compute pressures and internal energies from
  first-principles at all temperatures and provide a coherent equation
  of state. We compare our first-principles calculations with analytic
  equations of state, which tend to agree for temperatures above
  8$\times$10$^6$~K. Pair-correlation functions and the electronic
  density of states reveal an evolving plasma structure and ionization
  process that is driven by temperature and density. As we increase
  the density at constant temperature, we find that the ionization
  fraction of the 1s state decreases while the other electronic states
  move towards the continuum.  Finally, the computed shock Hugoniot
  curves show an increase in compression as the first and second
  shells are ionized.
\end{abstract}



\maketitle

\section{INTRODUCTION}

Elemental oxygen is involved in a wide range of physics and chemistry
throughout the universe, spanning from ambient biological processes to
extreme geological and astrophysical processes. Created during stellar
nucleosynthesis, oxygen is the third most abundant element in the
universe and the most abundant element on Earth. In addition to its
importance for life-sustaining processes, its thermodynamic, physical,
and chemical properties are important to numerous fields of
science. As such, oxygen has inspired a vast number of laboratory
experiments and theoretical studies, which have revealed an exotic
phase diagram with a number of interesting anomalies in its thermal,
optical, magnetic, electrical, and acoustic properties due to its
molecular and magnetic nature~\cite{Freiman2004}.

At ambient conditions, oxygen exists as a diatomic molecular gas with
each molecule having two unpaired electrons, resulting in a
paramagnetic state. X-ray diffraction and optical experiments reveal
that oxygen condenses to a molecular solid with a rich phase diagram
made up of at least ten different structural
phases~\cite{Freiman2004,Santoro2004,Lundegaard2006,Goncharov2011,Ma2007,Sun2012}. Static
compression experiments on the solid have been performed up to 1.3
Mbar and 650 K~\cite{Freiman2004}. First-principles simulations have
been used to search for structural phases up to 100
Mbar~\cite{Sun2012}. The transition to the highest-pressure phase
discovered so far occurs at 96 GPa, which also drives the solid to
become
metallic~\cite{Desgreniers1990,Akahama1995,Serra1998,Militzer2006}. A
superconducting phase has also been found at 0.6 K near 100
GPa~\cite{Shimizu1998}.  In addition, the solid phases exhibit a
complex magnetic structure with various degrees of ordering due to a
strong exchange interaction between O$_2$ molecules that becomes
suppressed under pressure and acts in tandem with weak van der Waals
forces holding the lattice together~\cite{Neaton2002,Freiman2004,
  Klotz2010}.

\begin{figure}[t]
  \begin{center}
        \includegraphics*[width=8.6cm]{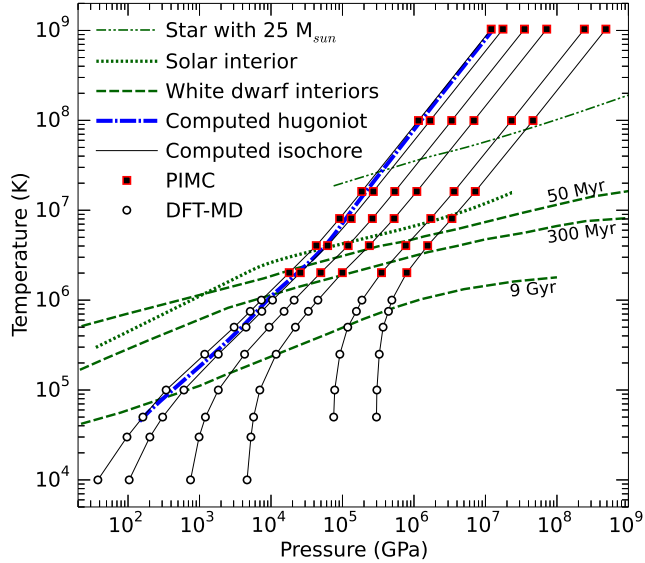}
  \end{center}

    \caption{Temperature-pressure conditions for the PIMC and DFT-MD
      calculations along six isochores corresponding to the densities
      of 2.48634, 3.63046, 7.26176, and 14.8632, 50.00, and 100.00
      g$\,$cm$^{-3}$. The dash-dotted line shows the Hugoniot curve
      for an initial density of $\rho_0 = 0.6671$ g$\,$cm$^{-3}$. For
      comparison, we also plotted the interior profile of the
      current-day Sun\cite{Bahcall2004} as well as the profile of a 25
      M$_{\odot}$ star at the end of its helium burning time
      \cite{GeorgyPC}. The green dashed lines show the interior
      profile of a 0.6 M$_{\odot}$ carbon-rich white dwarf at three
      different stages of its cooling process
      \cite{Fontaine1976,Fontaine1977,Hansen2004}}

  \label{fig:HugTvsP}
\end{figure}

Warm, dense, fluid states of oxygen have also been of great interest
due to the presence of oxygen-rich compounds in inner layers of giant
planets~\cite{Burrows2001, Hubbard2002, Guillot2005, Wilson2012,
  Zhang2013}, stellar interiors~\cite{Hansen1994, Fortney2012},
astrophysical processes~\cite{Lodders2002, Zoccali2006, Wong2008}, and
detonation products~\cite{Needham2010}. Oxygen is produced via helium
burning~\cite{Wallerstein1997} in the late stages of Sun-like star's
life as well as in more massive stars. The larger weight of oxygen
relative to hydrogen and helium drives its settlement towards the
deepest regions of a star. An accurate equation of state (EOS) is
needed to properly describe the behavior of the core of the star as
well as the timing of the different nuclear processes that are highly
sensitive to
temperature~\cite{Wallerstein1997,Burbidge1957}. Eventually,
intermediate mass stars evolve into white dwarfs, which have most of
their hydrogen and helium depleted, leaving a remnant composed mostly
of carbon and oxygen. The core density of a white
dwarf~\cite{Fontaine1976} is likely higher than 10$^5$~g/cm$^3$. The
cooling process of the white dwarf is very similar from one white
dwarf to another and the luminosity is used for cosmological
chronology~\cite{Schmidt1959,Fontaine2001}. However, the accuracy of
chronology measurements depends on a proper description of the
thermodynamic behavior of both carbon and
oxygen~\cite{Chabrier2000}. Moreover, as the third most abundant
element in the solar system~\cite{Lodders2003}, oxygen has a
significant presence in planet interiors and can exist in a partially
ionized state in giant planets. Therefore, the electronic and
thermodynamic behavior of oxygen at high pressures and temperatures is
important for obtaining the correct fluid and magnetic behavior in
planetary, stellar, and stellar remnant models~\cite{Itoh1990}.

Shock-compressed fluid states of oxygen have been measured under
dynamic compression up to 1.9 Mbar (four-fold compression) and 7000 K,
which revealed a metallic transition in the molecular fluid at 1.2
Mbar and 4500 K~\cite{Bastea2001}. Density functional theory molecular
dynamics (DFT-MD) simulations suggest that disorder in the fluid
lowers the metallization pressure to as low as 30 GPa with molecular
dissociation above 80 GPa~\cite{Militzer2003}. Measurements of
Hugoniots have reached 140
GPa~\cite{Nellis1980,Hamilton1988,Chisolm2009} and indicate that
oxygen molecules become dissociated in a pressure range of 80-120 GPa
at temperatures over several thousand Kelvin. Using classical
pair-potential simulations~\cite{Ross1980,Kerley1986,Chen2008}, some
general agreement is found with the measured Hugoniots, however, a
fully quantum-mechanical treatment is needed to accurately simulate
the electronic and structural behavior of the fluid.

Historically, a lack of development in first-principles methodology
for the warm dense matter regime has largely prevented highly accurate
theoretical exploration of fluid oxygen at extreme conditions, and,
hence, further improvements in EOS and Hugoniot curves. DFT-MD has
been used to explore the structural and electronic behavior of the
fluid state~\cite{Wang2010, Militzer2003} up to temperatures of
16$\times$10$^3$ K and densities up to 4.5 g$\,$cm$^{-3}$. Massacrier
\textit{et al.} \cite{Massacrier2011} investigated the properties of
oxygen for a density-temperature range of
$10^{-3}-10^4$~g$\,$cm$^{-3}$ and $10^5-10^6$~K, using an average ion
model. They showed, for instance, that the complete
pressure-ionization of fluid oxygen cannot be expected until the
system reaches a density of 1000~g$\,$cm$^{-3}$.

In order to address the challenges of first-principles simulations for
warm dense matter, we have been developing the path integral Monte
Carlo (PIMC) methodology in recent years for the study of heavy
elements in warm, dense states~\cite{Mi09, Driver2012,
  BenedictCarbon, Driver2015}. Here, we apply our PIMC methodology
along with DFT-MD to extend the first principles exploration of warm
dense fluid oxygen to a much wider density-temperature range (1--100
g$\,$cm$^{-3}$ and 10$^4$--10$^9$ K) than has been previously explored
by DFT-MD alone.

In Section II, we cover details of the PIMC and DFT-MD methodology
specific to our oxygen simulations. In Section III, we discuss the EOS
constructed from PIMC and DFT-MD and show that both methods agree for
at least one of temperature in the range of
2.5$\times$10$^5$--1$\times$10$^6$ K. In section IV, we characterize
the structure of the plasma and the ionization process by examining
pair-correlation functions of electrons and nuclei as a function of
temperature and density. In section V, we discuss the electronic
density of states as a function of density and temperature to provide
further insight into the ionization process. In section VI, we discuss
predictions for the shock Hugoniot curves. Finally, in section VII, we
summarize and conclude our results.

\section{SIMULATION METHODS}

PIMC~\cite{Ceperley1995, Mi09} is currently the state-of-the-art
first-principles method for simulating materials at temperatures in
which properties are dominated by excited states. It is the only
method able to accurately treat all the effects of bonding,
ionization, exchange-correlation, and quantum degeneracy that
simultaneously occur in the warm dense matter
regime~\cite{Koenig2005}. PIMC is based on thermal density matrix
formalism, which is efficiently computed with Feynman's imaginary time
path integrals. The density matrix is the natural operator to use for
computing high-temperature observables because it explicitly includes
temperature in a many-body formalism.

The PIMC method stochastically solves the full, finite-temperature
quantum many-body problem by treating electrons and nuclei equally as
quantum paths that evolve in imaginary time without invoking the
Born-Oppenheimer approximation. For our PIMC simulations, the Coulomb
interaction is incorporated via pair density matrices derived from the
eigenstates of the two-body Coulomb problem~\cite{Pollock1988,
  Ceperley1995} appropriate for oxygen. Furthermore, in contrast to
DFT-MD as described below, the efficiency of PIMC increases with
temperature as particles behave more classical-like and fewer time
slices are needed to describe quantum mechanical many-body
correlations, scaling inversely with temperature.

\begin{figure}[t]
  \begin{center}
        \includegraphics*[width=8.6cm]{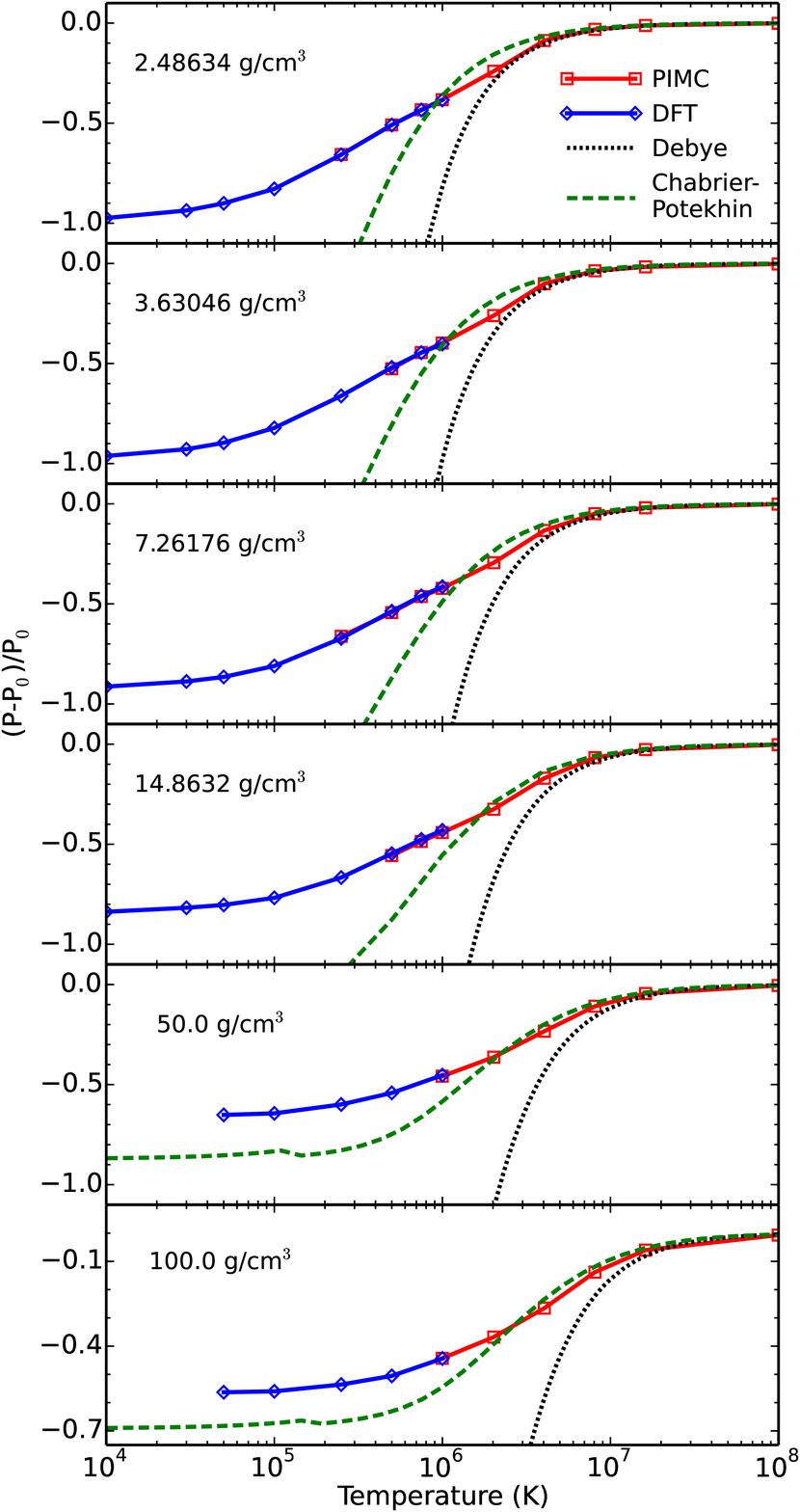}
  \end{center}

    \caption{Comparison of excess pressure relative to
      the ideal Fermi gas plotted as a function of temperature for
      oxygen.}

  \label{fig:PvsT}
\end{figure}

\begin{figure}[t]
  \begin{center}
        \includegraphics*[width=8.6cm]{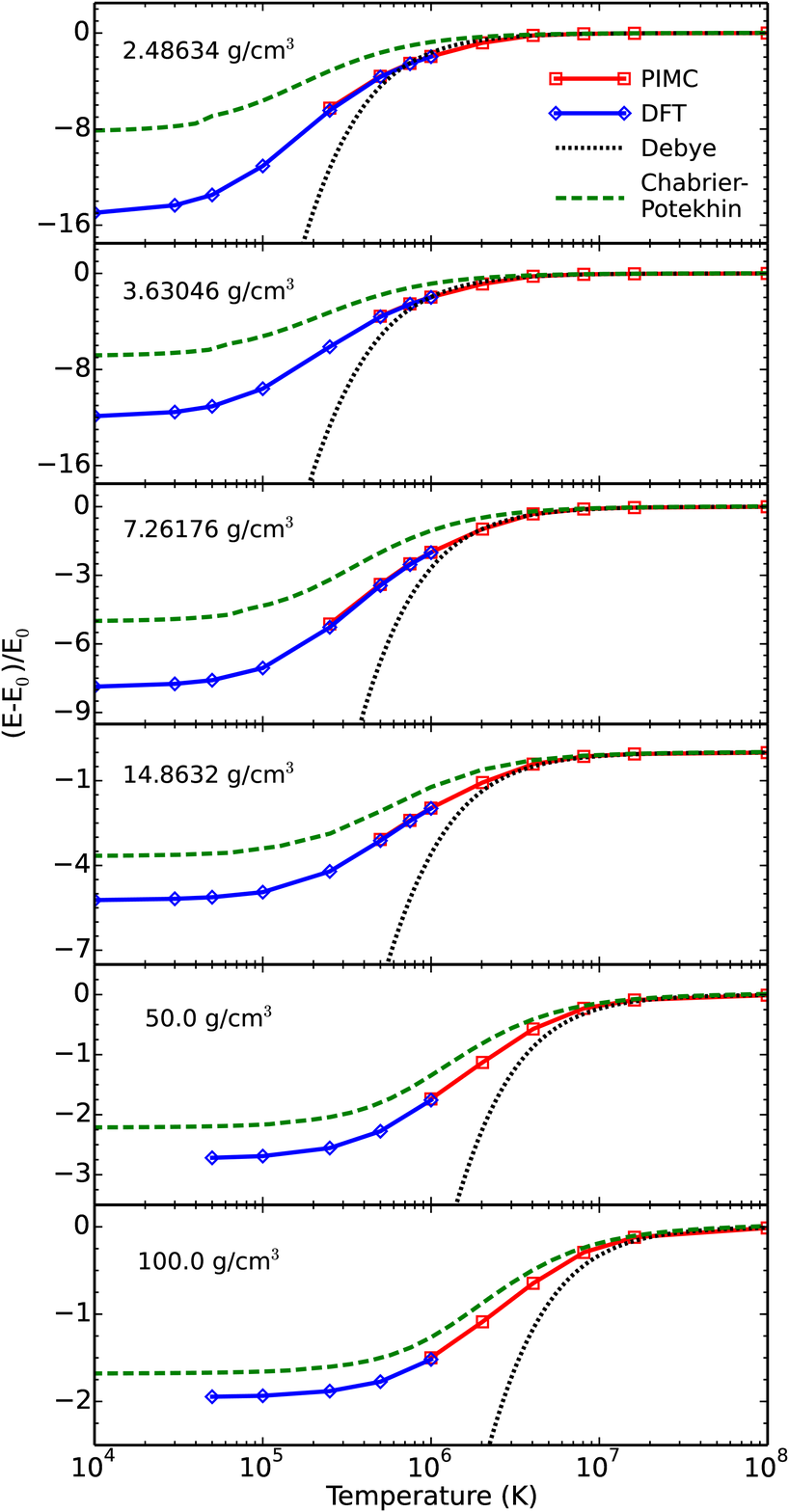}
  \end{center}

    \caption{Comparison of excess internal energies
      relative to the ideal Fermi gas plotted as a function of
      temperature for oxygen.}

  \label{fig:IEvsT}
\end{figure}

PIMC uses a minimal number of controlled approximations, which become
vanishingly small with increased temperature and by using appropriate
convergence of the time-step and system size. The only uncontrolled
approximation is the employment of a fixed nodal surface to avoid the
fermion sign problem~\cite{Ceperley1991}. Current state-of-the art
PIMC calculations employ a free-particle nodal structure, which would
perfectly describe a fully ionized system.  However, we have shown
PIMC employing free-particle nodes even produces reliable results at
surprising low temperatures in partially ionized
hydrogen~\cite{Militzer2001a}, carbon~\cite{Driver2012},
water~\cite{Driver2012}, and neon~\cite{Driver2015}.  As a general
rule, we find free-particle nodes are sufficient for systems comprised
of partially-ionized 2s states~\cite{Driver2012}.

A sufficiently small PIMC time step is determined by converging total
energy as a function of time step until the energy changes by less
than 0.5\%, which is shown in supplemental material~\cite{SupMat}
Table SI. We use a time step of 1/256 Ha$^{-1}$ for temperatures below
4$\times10^6$ K and, for higher temperatures, we decrease the time
step as $1/T$, as the efficiency of PIMC increases linearly with T as
path lengths decrease. The number of time slices we use in the path
integral range from 323 at lowest temperature to 5 at the highest
temperature. In order to minimize finite size errors, the internal
energy and pressure is converged to better than 0.4\% when comparing
8- and 24-atom simple cubic simulation cells, which is shown in
supplemental material~\cite{SupMat} Table SII. We therefore perform
all PIMC calculations in 8-atom cells, as PIMC scales as N$^2$, where
N is the number of particles. A typical calculation uses a bisection
level~\cite{Ceperley1995} of 5 and achieves a statistical error in the
energy and pressure that is less than 0.1\%.

For lower temperatures (T $<$ 1$\times$10$^6$ K),
DFT-MD~\cite{marx2000ab} is the most efficient state-of-the-art
first-principles method. DFT formalism provides an exact mapping of
the many body problem onto a single particle problem, but, in
practice, employs an approximate exchange-correlation potential to
describe many body electron physics. In the WDM regime, where
temperatures are at or above the Fermi temperature, the
exchange-correlation functional is not explicitly designed to
accurately describe the electronic physics~\cite{Brown2013}. However,
in previous PIMC and DFT-MD work on helium~\cite{Mi09}
carbon~\cite{Driver2012}, and water~\cite{Driver2012}, and
neon~\cite{Driver2015}, DFT functionals are shown to be accurate even
at high temperatures.

DFT incorporates effects of finite electronic temperature into
calculations by using a Fermi-Dirac function to allow for thermal
occupation of single-particle electronic states~\cite{Mermin1965}. As
temperature grows large, an increasing number of bands are required to
account for the increasing occupation of excited states in the
continuum, which typically causes the efficiency of the algorithm to
become intractable at temperatures beyond 1$\times$10$^6$
K. Orbital-free density functional methods aim to overcome such
thermal band efficiency limitations, but several challenges remain to
be solved~\cite{Lambert2006}. In addition, pseudopotentials, which
replace the core electrons in each atom and improve efficiency, may
break down at temperatures where core electrons undergo excitations.

Depending on the density, we employ two different sets of DFT-MD
simulations for our study of oxygen.  At densities below 15~g
cm$^{-3}$, the simulations were performed with the Vienna $Ab~initio$
Simulation Package (VASP)~\cite{VASP} using the projector
augmented-wave (PAW) method~\cite{PAW}. The VASP DFT-MD uses a NVT
ensemble regulated with a Nos\'{e}-Hoover
thermostat. Exchange-correlation effects are described using the
Perdew-Burke-Ernzerhof~\cite{PBE} generalized gradient
approximation. Electronic wave functions are expanded in a plane-wave
basis with a energy cut-off of at least 1000 eV in order to converge
total energy.  Size convergence tests up to a 24-atom simulation cell
at temperatures of 10,000 K and above indicate that total energies are
converged to better than 0.1\% in a 24-atom simple cubic cell. We
find, at temperatures above 250,000 K, 8-atom supercell results are
sufficient since the kinetic energy far outweighs the interaction
energy at such high temperatures. The number of bands in each
calculation is selected such that thermal occupation is converged to
better than 10$^{-4}$, which requires up to 8,000 bands in a 24-atom
cell at 1$\times$10$^6$ K. All simulations are performed at the
$\Gamma$-point of the Brillouin zone, which is sufficient for high
temperature fluids, converging total energy to better than 0.01\%
relative to a comparison with a grid of k-points.

For densities above 15~g cm$^{-3}$, we had to construct a new
pseudopotential in order to prevent the overlap of the PAW-spheres. We
therefore used the ABINIT package~\cite{Gonze2009} for which it is
possible to build a specific PAW-pseudopotential using the AtomPAW
plugin~\cite{Holzwarth2001}. We built a hard all-electron PAW
pseudopotential with a cut-off radius of 0.4 Bohr. We checked the
accuracy of the pseudopotential by reproducing the results provided by
the ELK software in the linearized augmented plane wave (LAPW)
framework~\cite{ELK}. With this pseudopotential we performed DFT-MD
with ABINIT for a 24-atom cell up to 100~g~cm$ ^{-3}$ and
1$\times$10$^6$~K. The hardness of the pseudopotential required an
plane-wave energy cut-off of at least 6800~eV.

\section{EQUATION OF STATE RESULTS}

In this section, we report our EOS results for six densities of
2.48634, 3.63046, 7.26176, and 14.8632, 50.00, and 100.00
g$\,$cm$^{-3}$ and for a temperature range of $10^4-10^9$ K. The six
isochores are shown in Figure~\ref{fig:HugTvsP} and are discussed in
more detail in section VI. These conditions are relevant for the
modeling of stars and white dwarfs as can be seen in
Figure~\ref{fig:HugTvsP}.

Figure~\ref{fig:PvsT} compares pressures obtained for oxygen from
PIMC, DFT-MD, and from analytic Chabrier-Potekhin~\cite{Chabrier1998}
and Debye-H\"{u}ckel~\cite{DebyeHuckel} models. Pressures, $P$, are
plotted relative to a fully ionized Fermi gas of electrons and ions
with pressure, $P_{0}$, in order to compare only the excess pressure
contributions that result from particle interactions. In general PIMC
and DFT-MD pressures differ by at most 2\%, and often much less for at
least one temperature in the range of $2.5\times10^5-1\times10^6$
K. PIMC converges to the weakly interacting plasma limit along with
the Chabrier-Potekhin and Debye-H\"{u}ckel models.

Figure~\ref{fig:IEvsT} compares internal energies, $E$, plotted
relative to the internal energy of a fully ionized Fermi gas,
E$_0$. PIMC and DFT-MD results for excess internal energy differ by at
most 2\%, and much less in most cases for at least one temperature in
the range of $2.5\times10^5-1\times10^6$ K. PIMC extends the energies
to the weakly interacting plasma limit at high temperatures, in
agreement with the Potekhin and Debye-H\"{u}ckel
models~\cite{DebyeHuckel}.

Together, Figs.~\ref{fig:PvsT} and ~\ref{fig:IEvsT} show that the
DFT-MD and PIMC methods form a coherent equation of state over all
temperatures ranging from the regime of warm dense matter to the
weakly interacting plasma limit. The agreement between PIMC and DFT-MD
indicates that DFT exchange-correlation potential remains valid even
at high temperatures and that the PIMC free-particle nodal
approximation is valid for a sufficient ionization fraction of the 2s
state. The analytic Chabrier-Potekhin and Debye-H\"{u}ckel models
agree with PIMC to temperatures as low as 8$\times$10$^6$~K. The
Debye-H\"{u}ckel model appears to have better agreement with PIMC at
low densities, while the Chabrier-Potekhin model agrees better with
PIMC at high densities. Neither analytic model includes bound states
and, therefore, cannot describe low temperature conditions.

Table~\ref{EOSTable} provides the densities, temperatures, pressures,
and energies used to construct our equation of state. The VASP DFT-MD
energies have been shifted by 74.9392 Ha/atom in order to bring the
PAW-PBE pseudpotential energy in alignment with all-electron energies
that we report with PIMC computations. The shift was calculated by
performing an all electron atomic calculation with the OPIUM
code~\cite{OPIUM} and a corresponding isolated-atom calculation in
VASP.

\begin{figure}[t]
  \begin{center}
        \includegraphics*[width=8.6cm]{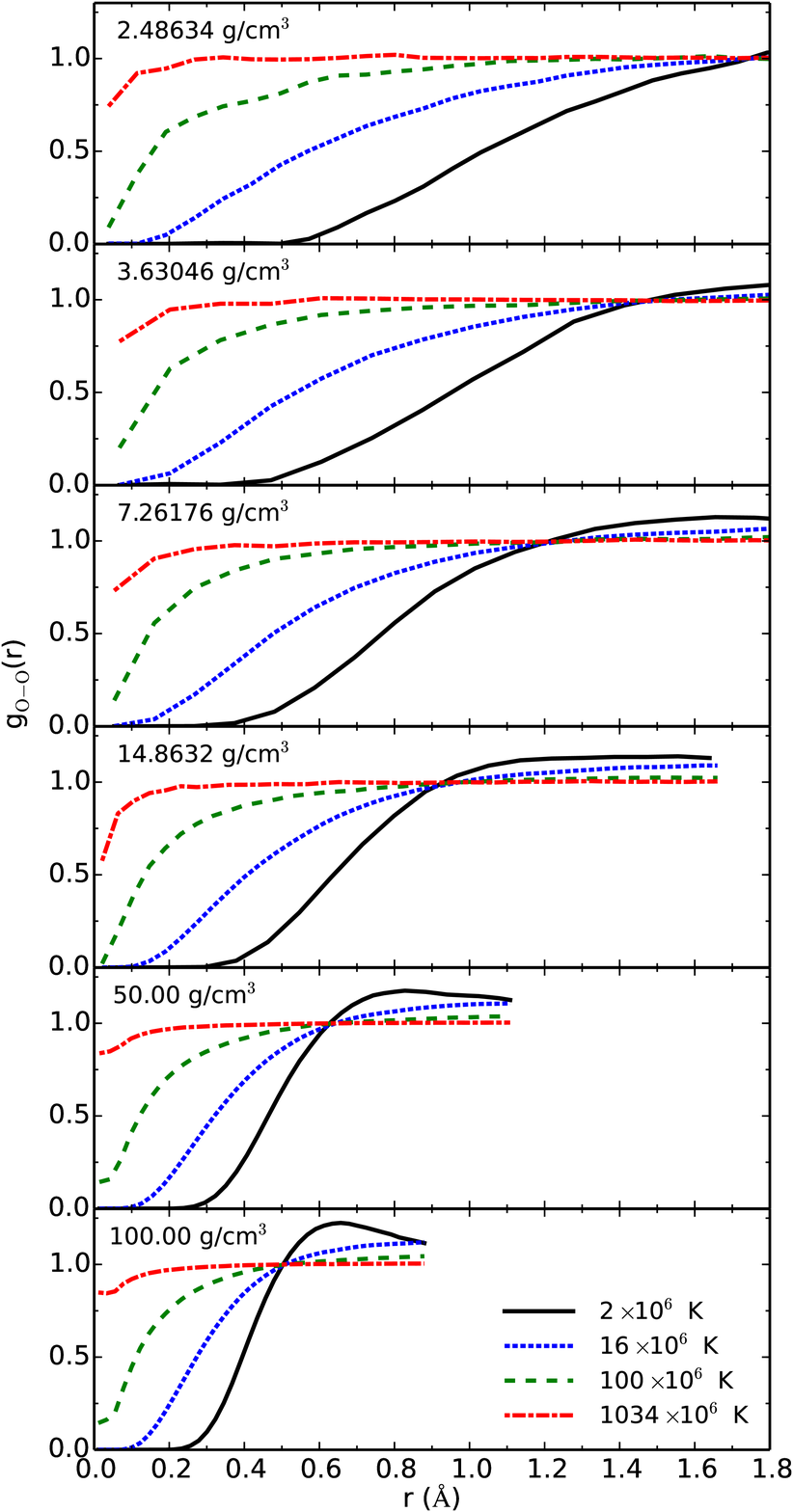}
  \end{center}

    \caption{Nuclear pair-correlation functions for oxygen from PIMC
      over a wide range of temperatures and densities.}

  \label{fig:gofroo}
\end{figure}

\begin{figure}[t]
  \begin{center}
        \includegraphics*[width=8.6cm]{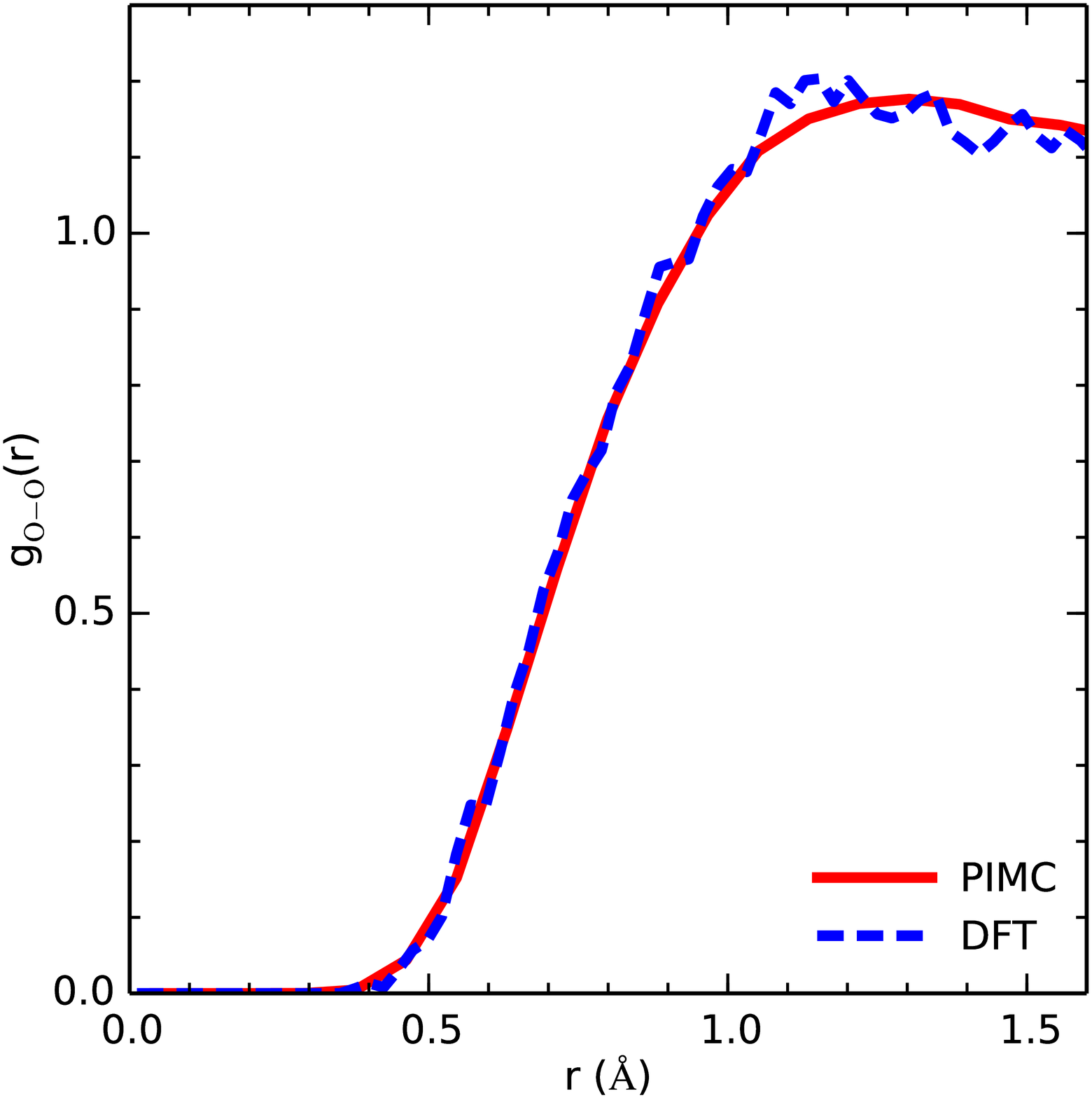}
  \end{center}

    \caption{Comparison of PIMC and DFT nuclear pair-correlation
      functions for oxygen at a temperature of 1$\times$10$^6$ K and a
      density of 14.8632 g$\,$cm$^{-3}$.}

  \label{fig:gofroo1millionK}
\end{figure}

Comparison of the PIMC and DFT-MD pressures and internal energies in
Table~\ref{EOSTable} indicates that there is roughly a 2\% discrepancy
in their predicted values at temperatures of $1\times10^6$
K. Potential sources of this discrepancy include: (1) the use of free
particle nodes in PIMC; (2) the exchange-correlation functional in
DFT; and (3) the use of a pseudopotential in DFT. While it is
difficult to determine the size of the nodal and exchange-correlation
errors, comparison of our VASP calculations with all-electron, PAW
ABINIT calculations at $1\times10^6$ K indicates that roughly one
third of the discrepancy is due to the use of frozen 1s core in the
VASP DFT-MD pseudopotential, which leaves out effects of core
excitations.

\section{PAIR-CORRELATION FUNCTIONS}

In this section, we study pair-correlation
functions~\cite{Militzer2009JPAM} in order to understand the evolution
of the fluid structure and ionization in oxygen plasmas as a function
of temperature and density.

Figure~\ref{fig:gofroo} shows the nuclear pair-correlation functions,
$g(r)$, computed with PIMC over a temperature range of
$2\times10^6-1.034\times10^{12}$ K and a density range of
$2.486-100.0$ g$\,$cm$^{-3}$. Atoms are kept farthest apart at low
temperatures due to a combination of Pauli exclusion among bound
electrons and Coulomb repulsion. As temperature increases, kinetic
energy of the nuclei increases, making it more likely to find atoms at
close range, and, in addition, the atoms become increasingly ionized,
which gradually minimizes the effects of Pauli repulsion. As density
increases, the likelihood of finding two nuclei at close range is
significantly increased. For the highest density and lowest
temperature, the peak in the pair-correlation function reaches a value
of 1.2, indicating a moderately structured fluid.

Figure~\ref{fig:gofroo1millionK} compares the nuclear
pair-correlation functions of PIMC and DFT at a temperature of
1$\times$10$^6$ K in an 8-atom cell at a density of 14.8632
g$\,$cm$^{-3}$. The overlapping $g(r)$ curves verify that PIMC and DFT
predict consistent structural properties.

\begin{figure}
  \begin{center}
        \includegraphics*[width=8.6cm]{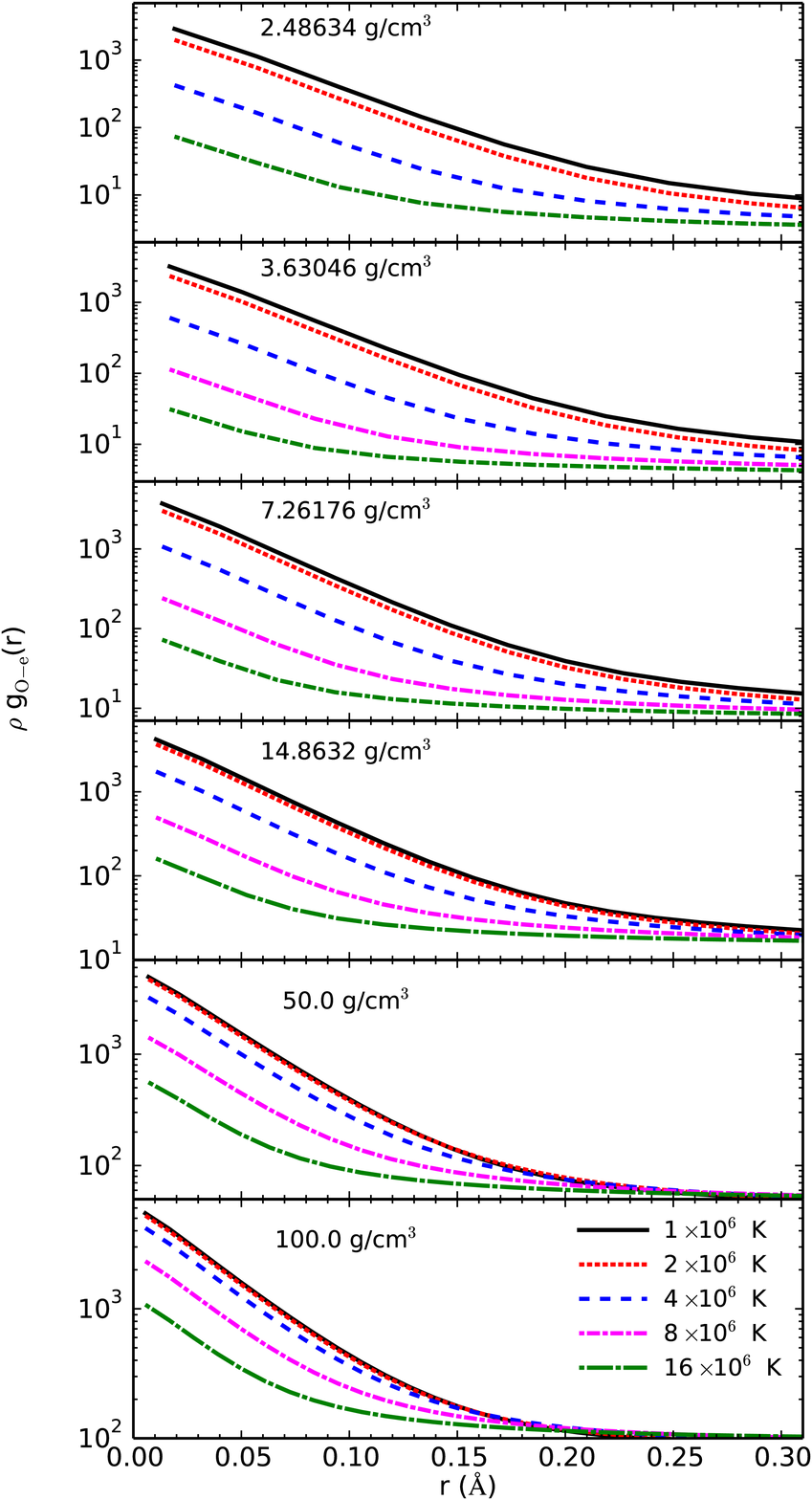}
  \end{center}

    \caption{The nucleus-electron pair-correlation functions for
      oxygen computed with PIMC.}

  \label{fig:gofroe}
\end{figure}

\begin{figure}
  \begin{center}
        \includegraphics*[width=8.6cm]{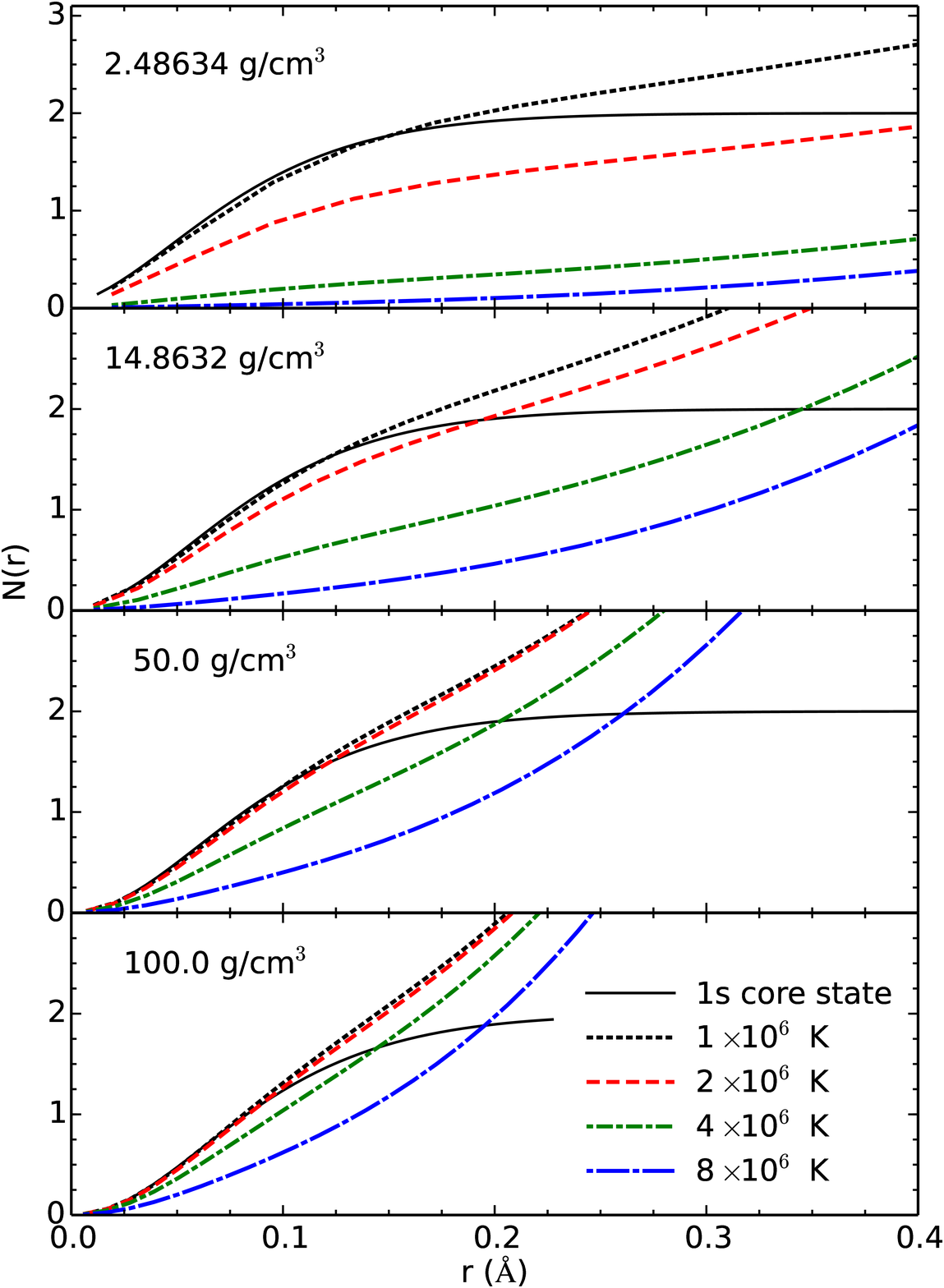}
  \end{center}

    \caption{Number of electrons contained in a sphere of radius, $r$,
      around an oxygen nucleus. PIMC data at four temperatures is
      compared with the analytic 1s core state.}

  \label{fig:nr}
\end{figure}

Figure~\ref{fig:gofroe} shows nucleus-electron pair correlation
functions.  Electrons are most highly correlated with the nuclei at
low temperature and high density, reflecting a lower ionization
fraction.  As temperature increases, electrons are thermally excited
and gradually become unbound, decreasing their correlation with the
nuclei. As the density is increased, the electrons are more likely to
reside near the nuclei, indicating that the ionization of
the 1s state is suppressed with increasing density.

Figure~\ref{fig:nr} shows the integral of the nucleus-electron pair correlation function, $N(r)$,
which represents the average number of electrons within a sphere of
radius $r$ around a given nucleus,
\begin{equation}
N(r) = \left< \frac{1}{N_I} \sum_{e,I} \theta(r-\left|\vec{r}_e-\vec{r}_I \right|) \right>,
\end{equation}
where the sum includes all electron-ion pairs and $theta$ represents
the Heaviside function. At the lowest temperature, 1$\times$10$^6$ K,
we find that the 1s core state is always fully occupied, as it agrees
closely with the result of an isolated 1s state. As temperature
increases, the atoms are gradually ionized and electrons become
unbound, causing $N(r)$ to decrease. As density increases, an
increasingly higher temperature is required to fully ionize the atoms,
confirming that the 1s ionization fraction decreases with density as
seen in Fig.~\ref{fig:gofroe}. The 1s state is thus not affected by
pressure ionization in the density range of consideration. As we will
explain the density of states section, the ionization of the 1s state
is suppressed because with increasing density, the Fermi energy
increases more rapidly than energy of the 1s state.

\begin{figure}
  \begin{center}
        \includegraphics*[width=8.6cm]{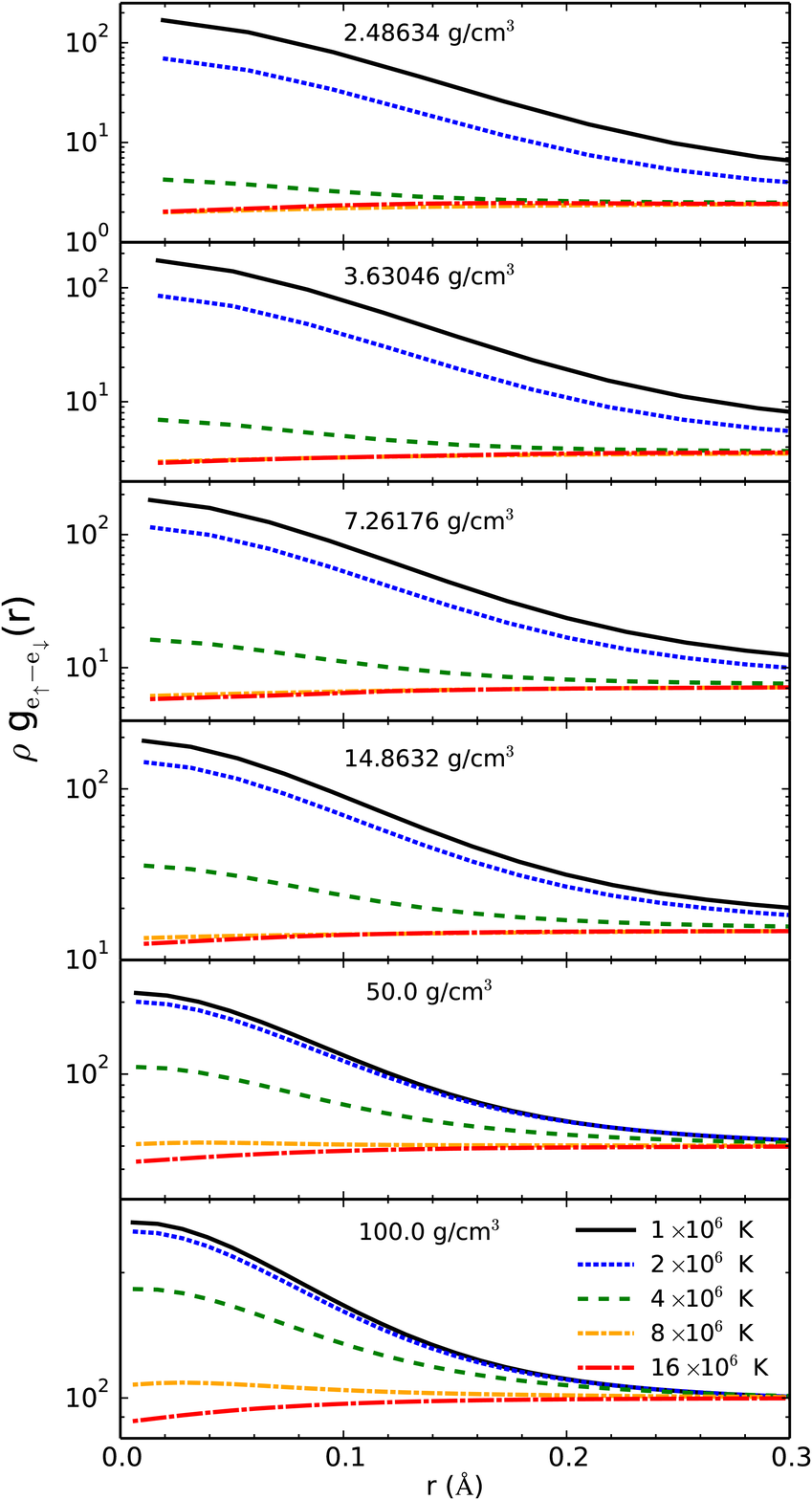}
  \end{center}

    \caption{The electron-electron pair-correlation functions
      (multiplied by $\rho$) for electrons with opposite spins
      computed with PIMC.}

  \label{fig:gofreeopp}
\end{figure}

\begin{figure}
  \begin{center}
        \includegraphics*[width=8.6cm]{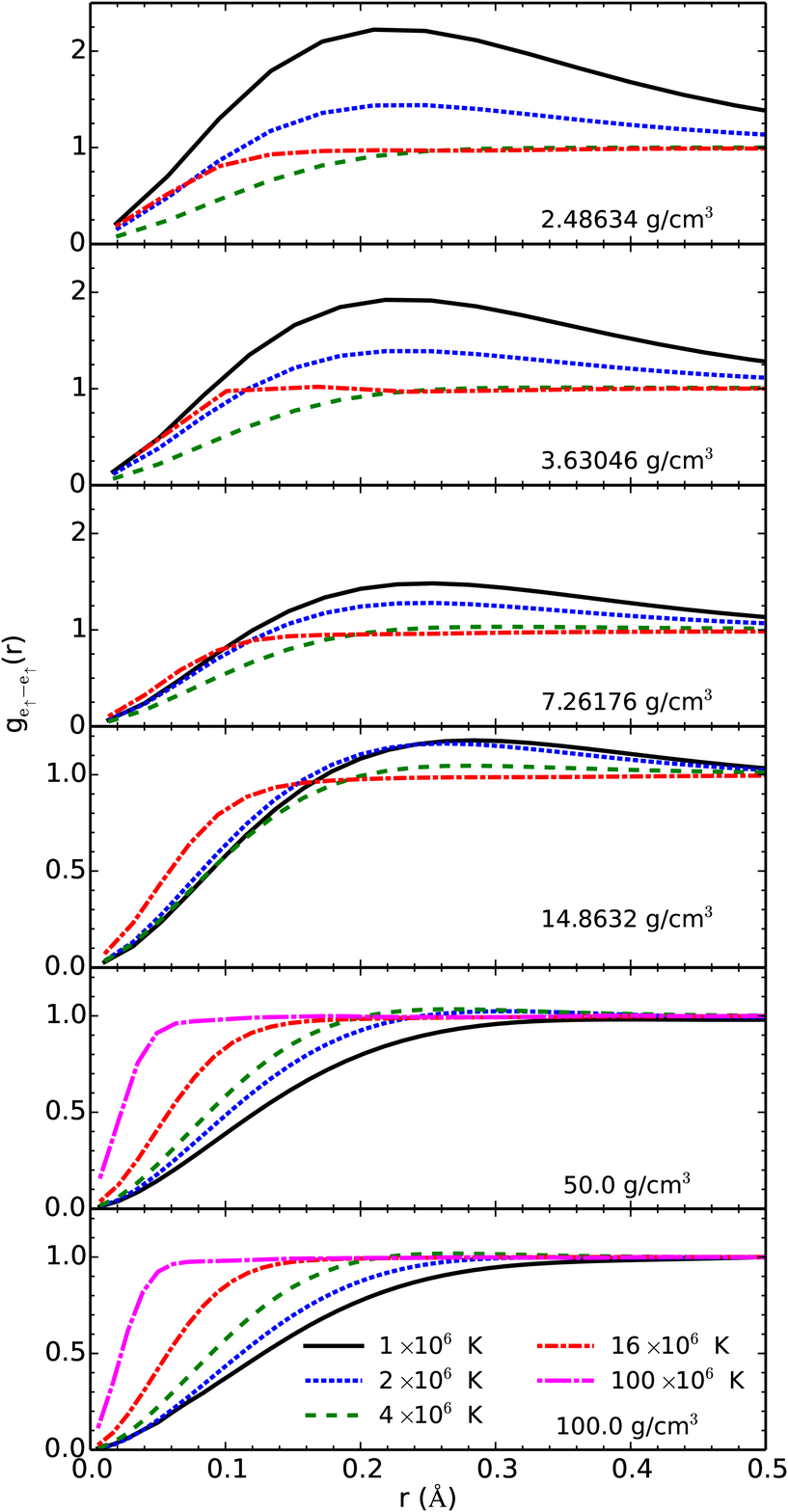}
  \end{center}

    \caption{The electron-electron pair-correlation functions for
      electrons with parallel spins computed with PIMC.}

  \label{fig:gofreepara}
\end{figure}

Figure~\ref{fig:gofreeopp} shows electron-electron pair correlations
for electrons having opposite spins. The function is multiplied by the
particle density, $\rho$, in units of g$~$cm$^{−3}$, so that the
integral under the curves is proportional to the number of
electrons. The electrons are most highly correlated for low
temperatures, which reflects that multiple electrons occupy bound
states at one nucleus.  As temperature increases, electrons are
thermally excited, decreasing the correlation among each
other. Correlation at short distances increases with density,
consistent with a lower ionization fraction.

Figure~\ref{fig:gofreepara} shows electron-electron pair correlations
for electrons with parallel spins. The positive correlation at at
$\sim$2.5 $\rm \AA$ for $T \le 2 \times 10^6\,$K reflects that
different electrons with parallel spins are bound to a single
nucleus. For short separations, Pauli exclusion takes over and the
functions decay to zero. The ordering of the g(r) curves changes with
respect to temperature as density increases due to a competition
between Coulomb and kinetic effects, coupled with the effects of
ionization. When the density is 50 and 100 g$\,$cm$^{-3}$, pressure
ionization causes the correlation to approach that of an ideal fluid,
and increasing temperature further only strengthens kinetic
effects. We interpret this change as pressure ionization of the second
and third electron shells. As temperature increases, electrons become
less bound, which also causes the correlation to become more like an
ideal fluid.

\section{ELECTRONIC DENSITY OF STATES}

In this section, we report DFT-MD results for the electronic density
of states (DOS) of fluid oxygen as a function of temperature and density in
order to gain further insight into the temperature- and
pressure-ionization.

\begin{figure}
  \begin{center}
        \includegraphics*[width=8.6cm]{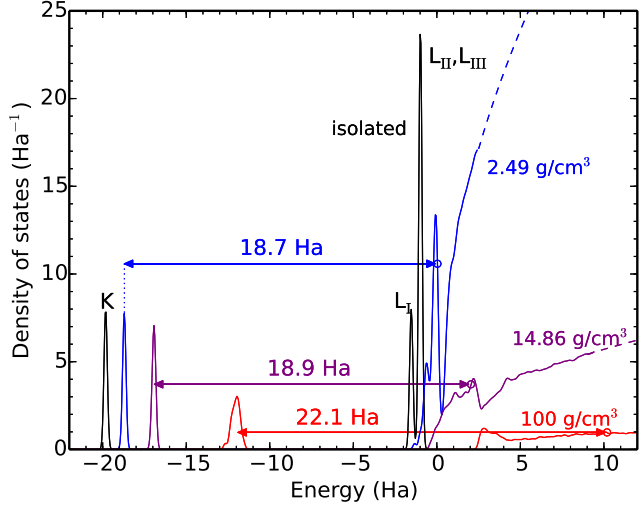}
  \end{center}

  \caption{Electronic density of states of dense, fluid oxygen using
    an all-electron, PAW pseudo-potential. The solid lines represent
    all available states for the isolated atom as well as three other
    densities at a temperature of 1$\times$10$^5$~K. The curves are
    normalized such that the occupied DOS integrates to 8. The K,
    L$_\textrm{I}$, L$_\textrm{II}$ and L$_\textrm{III}$ identify the
    electronic shells and sub-shells for the isolated atoms. The open
    circle on each curve stands for the DOS at the Fermi energy
    level. The arrows show the energy difference between the K-shell
    and the Fermi energy for the different densities.}

  \label{fig:DOSa}
\end{figure}

In order to closely examine the physics of pressure-ionization of the
1s and higher states, we computed DOS curves using the all-electron,
PAW potential we created for use with the ABINIT code.
Figure~\ref{fig:DOSa} shows examples of the DOS for oxygen at
densities between 2.49 and 100~g$\,$cm$^{-3}$ at a fixed temperature
of 100,000~K. For comparison, we show the result for an isolated
oxygen atom. Since we used the all-electron pseudo-potential we can
see the bands related to the 1s or K shell. For the isolated atom, we
also clearly see the 2s or L$_\textrm{I}$ as well as the
L$_\textrm{II}$ and L$_\textrm{III}$ states. The locations of the K
and L$_\textrm{I}$ shells for the isolated atom are consistent with
the binding energies of 19.97 and 1.53~Ha respectively that can be
found in the literature~\cite{Cardona1978}.

As density increases, the L sub-shells are shifted towards higher
energy, merging together as they shift into the continuum. This effect
is referred to as the pressure ionization of oxygen, also described by
Massacrier \textit{et al.}~\cite{Massacrier2011}. As the density
increases, the K shell is also shifted to higher energies and broadens
significantly. Nevertheless, the K shell remains a well defined state
even at 100~g$\,$cm$^{-3}$. The Fermi energy is also shifted towards
higher energy values as the density increases. We observe that the
Fermi energy shifts more than the K-shell energy, and, hence, the
energy difference between the 1s states and unoccupied states
increases with the density. Therefore, it is more difficult to
temperature-ionize the K shell at higher density and no
pressure-ionization occurs for the 1s state. This is consistent with
the observations we made for the electron-nuclei pair distribution
function in Fig.~\ref{fig:gofroe}.

\begin{figure}
  \begin{center}
        \includegraphics*[width=8.6cm]{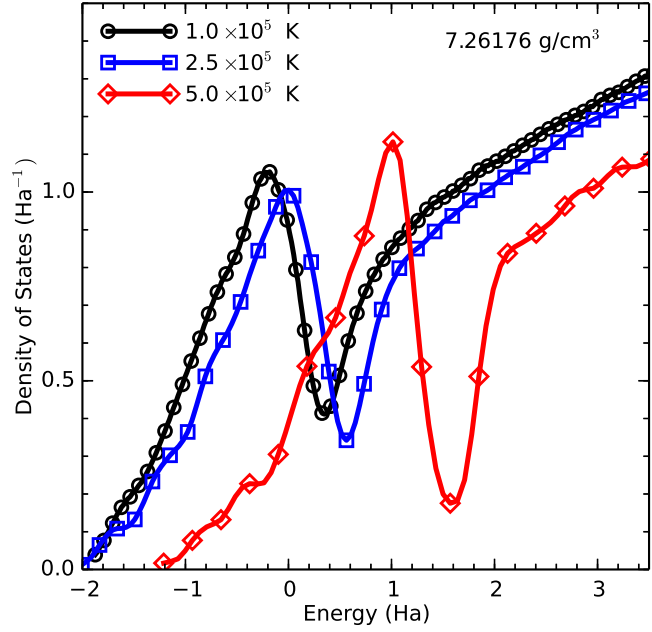}
  \end{center}

  \caption{Total electronic DOS of dense, fluid oxygen at a fixed
    density of 7.26176 g$\,$cm$^{-3}$ for three temperatures
    (1$\times$10$^5$, 2.5$\times$10$^5$ and 5$\times$10$^5$ K). Each
    DOS curve has had the relevant Fermi energy for each temperature
    subtracted from it.}

  \label{fig:DOSb}
\end{figure}

Figure~\ref{fig:DOSb} shows the temperature dependence of the DOS at a
fixed density of 7.26176 g$\,$cm$^{-3}$. Results were obtained from
VASP by averaging over at least 10 uncorrelated snapshots chosen from
a DFT-MD trajectory. Smooth curves were obtained by using a
4$\times$4$\times$4 k-point grid and applying a Gaussian smearing of
2~eV. The eigenvalues of each snapshot were shifted so that the Fermi
energies align at zero, and the integral of the DOS is normalized to
1. The DOS curves show a large peak representing the atomic-like 2s
and 2p states, followed by a dip in states, which is then followed by
a continuous spectrum of conducting states.  The Fermi energy plays
the role of the chemical potential in the Fermi-Dirac distribution,
which shifts towards more negative values as the temperature is
increased. Because we subtract the Fermi energy from the eigenvalues,
the peak shifts to higher energies with increasing temperature. The
fact that the peaks are embedded into a dense, continuous spectrum of
eigenvalues indicates that they are conducting states.

\section{SHOCK COMPRESSION}

\begin{figure}
  \begin{center}
        \includegraphics*[width=8.6cm]{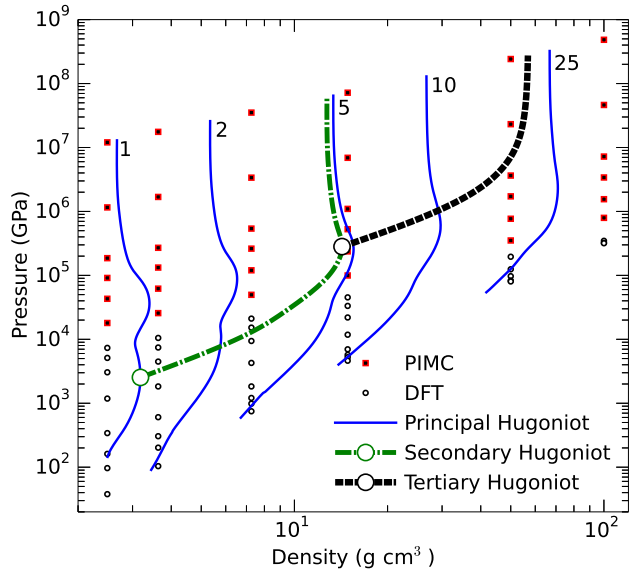}
  \end{center}

  \caption{Shock Hugoniot curves for different initial densities. The
    label on the curve specifies the ratio of the initial density to
    that of solid oxygen at 0K, 0.6671 g$\,$cm$^{-3}$. Secondary and
    tertiary Hugoniot curves are also plotted.}

  \label{fig:hugoniot1}
\end{figure}

\begin{figure}
  \begin{center}
        \includegraphics*[width=8.6cm]{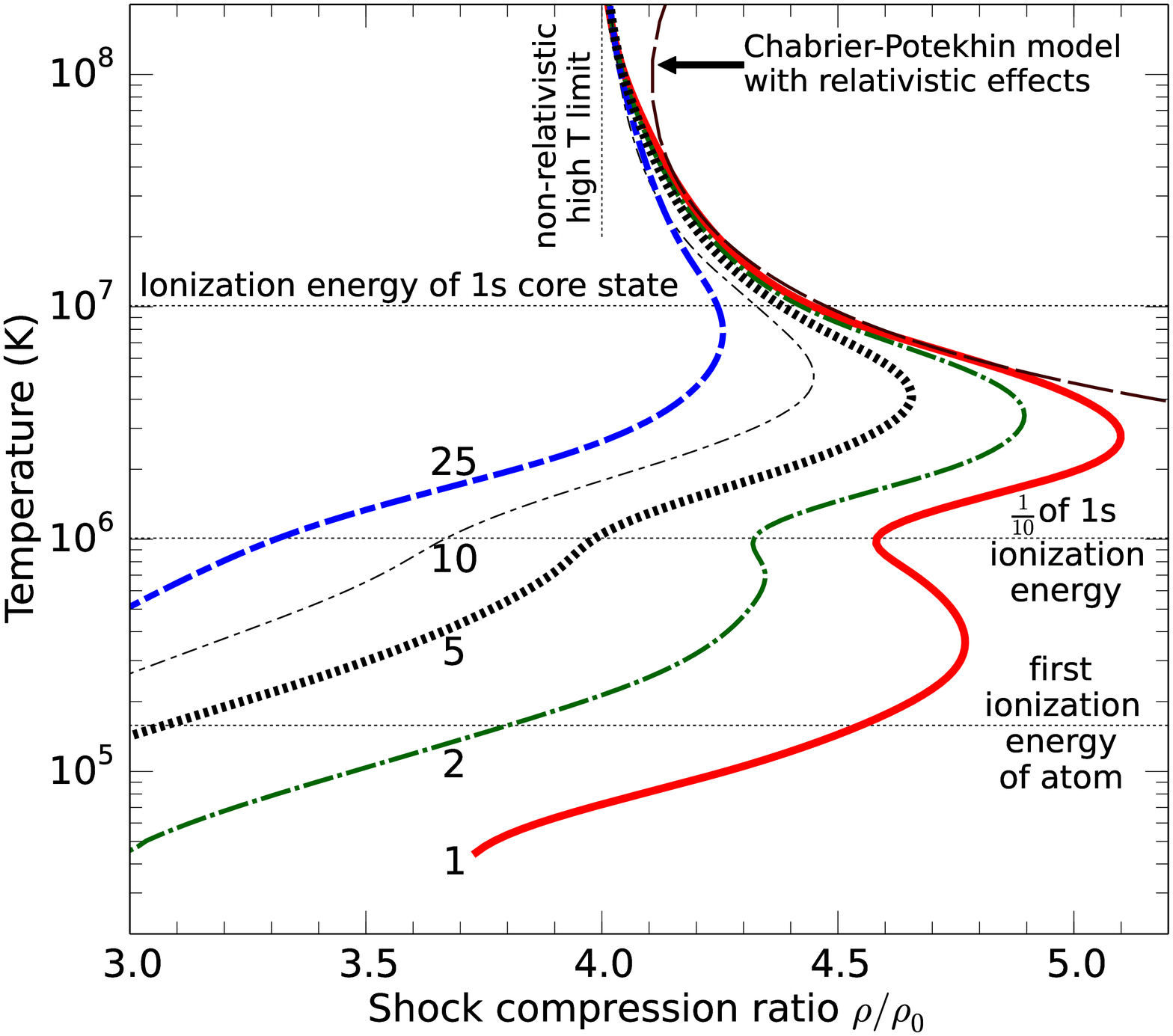}
  \end{center}

    \caption{Hugoniot curves for different pre-compression density
      ratios.}

  \label{fig:hugoniot2}
\end{figure}

Dynamic shock compression experiments are widely used for measuring
equation of state and other physical properties of hot, dense
fluids. Commonly, shock experiments determine the Hugoniot, which is
the locus of final states that can be obtained from different shock
velocities. A few Hugoniot measurements have been made for oxygen in
an effort to understand its metallic transition and determine its role
in astrophysical
processes~\cite{Nellis1980,Hamilton1988,Chisolm2009}. Density
functional theory has been validated by experiments as an accurate
tool for predicting the shock compression of different
materials~\cite{Root2010,Wang2010}.

In the course of a shock wave experiment, a material whose initial
state is characterized by an internal energy, pressure, and volume,
($E_0,P_0,V_0$), which changes to a final state denoted by $(E,P,V)$
while conserving mass, momentum, and energy. This leads to the
Rankine-Hugoniot relation~\cite{Ze66},
\begin{equation}
H = (E-E_0) + \frac{1}{2} (P+P_0)(V-V_0) = 0.
\label{hug}
\end{equation}

Here, we compute the Hugoniot for oxygen from the first-principles EOS
data we showed in Table~\ref{EOSTable}. The pressure and internal
energy data points were interpolated with bi-cubic spline functions in
$\rho-T$ space. For the initial state of the principal Hugoniot curve,
we computed the energy of an oxygen molecule at P$_0$ = 0, E$_0$ =
$-$150.247327 Ha/O$_2$, and chose V$_0$ = 318.612 \AA$^3$. We chose a
density of 0.6671 g$\,$cm$^{-3}$ for solid oxygen in the cubic,
$\gamma$ phase. The resulting Hugoniot curve has been plotted in
$T$-$P$ and $P$-$\rho$ spaces in Figs.~\ref{fig:HugTvsP}
and~\ref{fig:hugoniot1}, respectively.

Samples in shock wave experiments may be pre-compressed inside of a
diamond anvil cell in order to reach much higher final densities than
possible with a sample at ambient conditions. This technique allows
shock wave experiments to probe density-temperature consistent with
planetary and stellar interiors~\cite{MH08}. Therefore, we repeat our
Hugoniot calculation starting with initial densities ranging from a 1
to a 25-fold increase of the ambient density.
Figure~\ref{fig:hugoniot1} shows the resulting family of Hugoniot
curves. While starting from the ambient density leads to a maximum
shock density of 3.5 g$\,$cm$^{-3}$, a 25-fold pre-compression yields
a much higher maximum shock density of 71 g$\,$cm$^{-3}$, as expected.
However, such extreme densities can be reached more easily with triple
shock experiments as our example in Fig.~\ref{fig:hugoniot1}
illustrates. We used the first compression maximum on the principal
Hugoniot curve ($\rho=$3.182 g$\,$cm$^{-3}$, $P=$2535 GPa, $T=$
358,600 K) as the initial state of the secondary Hugoniot curve. The
compression maximum on this curve ($\rho=$14.25 g$\,$cm$^{-3}$,
$P=$282000 GPa, $T=$ 4,819,000 K) served as initial state for the
tertiary Hugoniot curve.

Figure~\ref{fig:hugoniot2} shows the temperature dependence of the
precompression density ratio for the five representative Hugoniot
curves in Figure~\ref{fig:hugoniot1}. In the high-temperature limit, all
curves converge to a compression ratio of 4, which is the value of a
nonrelativistic ideal gas. We also include of the Hugoniot curve
computed with the relativistic, fully-ionized Chabrier-Potekhin model,
which shows the relativistic correction in the high-temperature
limit. In general, the shock compression is determined by the
excitation of internal degrees of freedom, which increases the
compression, and interaction effects, which decrease the
compression~\cite{Mi06}. Consistent with our results for hydrogen,
helium~\cite{Mi09}, and neon~\cite{Driver2015} we find that an
increase in the initial density leads to a slight reduction in the
shock compression (Figure~\ref{fig:hugoniot2}) because particles
interact more strongly at higher density.

The shock-compression ratio also exhibits two maxima as a function of
temperature, which can be attributed to the ionization of electrons in
the first and second shell.  On the principal Hugoniot curve, the
first maximum of $\rho/\rho_0$=4.77 occurs at temperature of
$3.59\times10^5$ K (30.94 eV), which is above the first ionization
energy of the oxygen atom, 13.61 eV, but less than the second
ionization energy, 35.12 eV. A second compression maximum of
$\rho/\rho_0$=5.10 is found for a temperature of $2.87\times10^6$ K
(247.32 eV), which can be attributed to the ionization of the 1s core
states of the oxygen ions. The 1s ionization energy is 871.41 eV. This
is consistent with the ionization process we observe in
Figure~\ref{fig:nr}, where charge density around the nuclei is reduced
over the range of $2-8\times10^6$ K. Since DFT-MD simulations, which
use pseudopotentials to replace core electrons, cannot access physics
about core ionization, PIMC is a necessary tool to determine the
maximum compression along the principle Hugoniot curve.

\section{CONCLUSIONS}

In this work, we have combined PIMC with DFT-MD to construct a
coherent EOS for oxygen over wide range of densities and temperatures
that includes warm dense matter and plasmas in stars and stellar
remnants. The two methods validate each other in temperature range of
2.5$\times$10$^5$--1$\times$10$^6$ K, where both yield consistent
results. We compared our equation of state at high temperature with
the analytic models of Chabrier-Potekhin and Debye-H\"{u}ckel. The
deviations that we identified underline the importance for new methods
like PIMC to be developed for the study of warm dense matter. Nuclear
and electronic pair-correlations reveal a temperature- and
pressure-driven ionization process, where temperature-ionization of
the 1s state is suppressed while other states are efficiently ionized
as density increases up to 100 g$\,$cm$^{-3}$. Changes in the density
of states confirms the temperature- and pressure-ionization behavior
observed in the pair-correlation data.  Lastly, we find the ionization
imprints a signature on the shock Hugoniot curves and that PIMC
simulations are necessary to determine the state of the highest shock
compression. Our and Hugoniot and equation of state will help to build
more accurate models for stars and stellar remnants.

\begin{acknowledgments}
  
  We are grateful to Alexander Potekhin for helpful discussions about
  the fully ionized EOS. We are also grateful to Cyril Georgy for
  sharing his data and knowledge of massive star evolution. We are
  also grateful to Jan Vorberger for discussions on continuum
  lowering. This research is supported by the U. S. Department of
  Energy, grant DE-SC0010517. Computational support was provided by
  NERSC, NASA, and the Janus supercomputer, which is supported by the
  National Science Foundation (Grant No. CNS-0821794), the University
  of Colorado, and the National Center for Atmospheric Research.

\end{acknowledgments}


%

\begin{table}

\caption{EOS table of oxygen pressures and internal energies at
  density-temperature conditions simulated in this work. The
  numbers in parentheses indicate the statistical uncertainties of the
  DFT-MD and PIMC simulations.}

\begin{tabularx}{8.6cm}{l@{\hskip 5mm} r@{\hskip 4mm} r@{\hskip 4mm} r}
\hline
\hline\\[-5pt]
$\rho$ (g$\,$cm$^{-3}$)    &  T (K)        & P (GPa)       & E (Ha/atom) \\[1pt]
\hline\\[-6pt]
2.48634$^a$ &  1034730000 &  12031695(879) &   44227(3)\\
2.48634$^a$ &  99497670   &  1155684(608)  &   4242(2)\\
2.48634$^a$ &  16167700   &  185881(73)    &   674.57(29)\\
2.48634$^a$ &  8083850    &  91166(21)     &   323.90(9)\\
2.48634$^a$ &  4041920    &  43037(12)     &   138.71(6)\\
2.48634$^a$ &  2020960    &  17999(15)     &   16.06(7)\\
2.48634$^a$ &  998004     &  7336(9)       &   $-41.43(4)$\\
2.48634$^b$ &  1000000    &  7339(6)       &   $-42.41(2)$\\
2.48634$^a$ &  748503     &  5118(11)      &   $-50.66(4)$\\
2.48634$^b$ &  750000     &  5119(5)       &   $-51.84(18)$\\
2.48634$^a$ &  500000     &  3044(11)      &   $-59.30(4)$\\
2.48634$^b$ &  500000     &  3049(5)       &   $-60.58(3)$\\
2.48634$^a$ &  250000     &  1189(12)      &   $-66.94(5)$\\
2.48634$^b$ &  250000     &  1183(3)       &   $-69.293(3)$\\
2.48634$^b$ &  100000     &  341(1)        &   $-73.635(1)$\\
2.48634$^b$ &  50000      &  161(1)        &   $-74.571(1)$\\
2.48634$^b$ &  30000      &  97(1)         &   $-74.811(1)$\\
2.48634$^b$ &  10000      &  38(1)         &   $-75.015(1)$\\
\hline
\end{tabularx}
\label{EOSTable}
\end{table}

\begin{table}
TABLE I. {(\emph{Continued.})}\\
\begin{tabularx}{8.6cm}{l@{\hskip 4mm} r@{\hskip 3mm} r@{\hskip 3mm} r}
\hline
\hline\\[-5pt]
$\rho$ (g$\,$cm$^{-3}$)    &  T (K)        & P (GPa)       & E (Ha/atom) \\[1pt]
\\[-6pt]				
3.63046$^a$ &  1034730000 &  17566926(1904) &  44223(5)\\
3.63046$^a$ &  99497670   &  1685108(750)   &  4235(2)\\
3.63046$^a$ &  16167700   &  269993(107)    &  669.34(28)\\
3.63046$^a$ &  8083850    &  132427(35)     &  320.24(11)\\
3.63046$^a$ &  4041920    &  61955(18)      &  132.56(6)\\
3.63046$^a$ &  2020960    &  25689(28)      &  10.67(8)\\
3.63046$^a$ &  998004     &  10569(14)      &  $-42.93(4)$\\
3.63046$^b$ &  1000000    &  10507(14)      &  $-44.13(2)$\\
3.63046$^a$ &  748503     &  7433(14)       &  $-51.81(4)$\\
3.63046$^b$ &  750000     &  7443(8)        &  $-52.79(5)$\\
3.63046$^a$ &  500000     &  4414(15)       &  $-60.15(4)$\\
3.63046$^b$ &  500000     &  4483(5)        &  $-61.412(6)$\\
3.63046$^b$ &  250000     &  1831(3)        &  $-69.658(2)$\\
3.63046$^b$ &  100000     &  605(2)         &  $-73.686(2)$\\
3.63046$^b$ &  50000      &  305(1)1        &  $-74.565(1)$\\
3.63046$^b$ &  30000      &  202(2)         &  $-74.797(1)$\\
3.63046$^b$ &  10000      &  104(1)         &  $-74.992(1)$\\
\hline
\end{tabularx}
\label{EOSTable}
\end{table}

\begin{table}
TABLE I. {(\emph{Continued.})}\\
\begin{tabularx}{8.6cm}{l@{\hskip 4mm} r@{\hskip 3mm} r@{\hskip 3mm} r}
\hline
\hline\\[-5pt]
$\rho$ (g$\,$cm$^{-3}$)    &  T (K)        & P (GPa)       & E (Ha/atom) \\[1pt]
\\[-6pt]				
7.26176$^a$ &  1034730000 &  35142831(2985) &  44227(4)\\
7.26176$^a$ &  99497670   &  3374099(1777)  &  4237(2)\\
7.26176$^a$ &  16167700   &  538734(172)    &  664.43(26)\\
7.26176$^a$ &  8083850    &  261808(75)     &  311.36(11)\\
7.26176$^a$ &  4041920    &  120041(34)     &  119.03(5)\\
7.26176$^a$ &  2020960    &  49637(51)      &  1.74(7)\\
7.26176$^a$ &  998004     &  20964(31)      &  $-45.53(4)$\\
7.26176$^b$ &  1000000    &  21301(20)      &  $-46.16(4)$\\
7.26176$^a$ &  748503     &  15122(42)      &  $-53.17(5)$\\
7.26176$^b$ &  750000     &  15236(21)      &  $-54.51(2)$\\
7.26176$^a$ &  500000     &  9262(24)       &  $-61.27(3)$\\
7.26176$^b$ &  500000     &  9424(10)       &  $-62.652(7)$\\
7.26176$^a$ &  250000     &  4405(44)       &  $-67.78(5)$\\
7.26176$^b$ &  250000     &  4268(5)        &  $-70.098(2)$\\
7.26176$^b$ &  100000     &  1831(3)        &  $-73.613(2)$\\
7.26176$^b$ &  50000      &  1210(2)        &  $-74.382(2)$\\
7.26176$^b$ &  30000      &  986(5)         &  $-74.606(2)$\\
7.26176$^b$ &  10000      &  749(1)         &  $-74.813(1)$\\
\hline
\end{tabularx}
\label{EOSTable}
\end{table}

\begin{table}
TABLE I. {(\emph{Continued.})}\\
\begin{tabularx}{8.6cm}{l@{\hskip 4mm} r@{\hskip 3mm} r@{\hskip 3mm} r}
\hline
\hline\\[-5pt]
$\rho$ (g$\,$cm$^{-3}$)    &  T (K)        & P (GPa)       & E (Ha/atom) \\[1pt]
\\[-6pt]				
14.8632$^a$ &  1034730000 &  71917073(5787) &  44217(4)\\
14.8632$^a$ &  99497670   &  6899765(3226)  &  4230(2)\\
14.8632$^a$ &  16167700   &  1096035(364)   &  655.35(24)\\
14.8632$^a$ &  8083850    &  527445(141)    &  299.20(10)\\
14.8632$^a$ &  4041920    &  237350(67)     &  103.41(5)\\
14.8632$^a$ &  2020960    &  99599(98)      &  $-5.97(6)$\\
14.8632$^a$ &  998004     &  44297(52)      &  $-47.32(3)$\\
14.8632$^b$ &  1000000    &  45274(64)      &  $-47.95(4)$\\
14.8632$^a$ &  748503     &  32595(59)      &  $-54.80(3)$\\
14.8632$^b$ &  750000     &  33293(69)      &  $-55.76(4)$\\
14.8632$^a$ &  500000     &  21447(56)      &  $-61.86(3)$\\
14.8632$^b$ &  500000     &  21945(35)      &  $-63.21(1)$\\
14.8632$^b$ &  250000     &  11803(11)      &  $-69.884(4)$\\
14.8632$^b$ &  100000     &  6975(7)        &  $-72.907(3)$\\
14.8632$^b$ &  50000      &  5705(6)        &  $-73.590(2)$\\
14.8632$^b$ &  30000      &  5239(4)        &  $-73.815(1)$\\
14.8632$^b$ &  10000      &  4626(8)        &  $-74.057(1)$\\
\hline
\end{tabularx}
\label{EOSTable}
\end{table}

\begin{table}
TABLE I. {(\emph{Continued.})}\\
\begin{tabularx}{8.6cm}{l@{\hskip 4mm} r@{\hskip 3mm} r@{\hskip 3mm} r}
\hline
\hline\\[-5pt]
$\rho$ (g$\,$cm$^{-3}$)    &  T (K)        & P (GPa)       & E (Ha/atom) \\[1pt]
\\[-6pt]				
50.0000$^a$ &  1034730000 &  241912168(8061) & 44208(1)\\
50.0000$^a$ &  99497670   &  23165568(7204)  & 4215(1)\\
50.0000$^a$ &  16167700   &  3638714(751)    & 633.85(14)\\
50.0000$^a$ &  8083850    &  1721016(318)    & 272.08(6)\\
50.0000$^a$ &  4041920    &  768044(164)     & 78.29(3)\\
50.0000$^a$ &  2020960    &  351315(214)     & $-13.11(4)$\\
50.0000$^a$ &  998004     &  185345(210)     & $-46.12(4)$\\
50.0000$^c$ &  1000000    &  187281(611)     & $-47.36(11)$\\
50.0000$^c$ &  500000     &  118441(752)     & $-60.27(11)$\\
50.0000$^c$ &  250000     &  91835(1078)     & $-65.16(15)$\\
50.0000$^c$ &  100000     &  77796(541)      & $-67.49(7)$\\
50.0000$^c$ &  50000      &  75320(609)      & $-67.90(8)$\\
\hline
\end{tabularx}
\label{EOSTable}
\end{table}

\begin{table}
TABLE I. {(\emph{Continued.})}\\
\begin{tabularx}{8.6cm}{l@{\hskip 4mm} r@{\hskip 3mm} r@{\hskip 3mm} r}
\hline
\hline\\[-5pt]
$\rho$ (g$\,$cm$^{-3}$)    &  T (K)        & P (GPa)       & E (Ha/atom) \\[1pt]
\\[-6pt]				
100.000$^a$ &  1034730000 &  483702750(18188) & 44193(2)\\
100.000$^a$ &  99497670   &  46258880(13163)  & 4201(1)\\
100.000$^a$ &  16167700   &  7213882(1458)    & 617.35(13)\\
100.000$^a$ &  8083850    &  3396956(706)     & 254.73(7)\\
100.000$^a$ &  4041920    &  1553594(378)     & 68.31(4)\\
100.000$^a$ &  2020960    &  793543(497)      & $-10.07(5)$\\
100.000$^a$ &  998004     &  490625(1050)     & $-40.28(10)$ \\
100.000$^c$ &  1000000    &  490505(1367)     & $-41.78(12)$\\
100.000$^c$ &  500000     &  369913(2987)     & $-52.88(24)$\\
100.000$^c$ &  250000     &  326893(1556)     & $-56.75(12)$\\
100.000$^c$ &  100000     &  302710(1091)     & $-58.79(8)$\\
100.000$^c$ &  50000      &  298808(1064)     & $-59.13(8)$\\
\hline
\hline
\end{tabularx}
\label{EoSTablec}
\vspace{-6pt}
\begin{flushleft}
$^a$PIMC \\$^b$VASP-MD \\$^c$ABINIT-MD with a small-core, PAW pseudopotentials\\
\end{flushleft}
\end{table}

\end{document}


\title{First-Principles Equation of State and Electronic Properties of Warm Dense Oxygen}

\author{K. P. Driver}
 \affiliation{Department of Earth and Planetary Science, University of California, Berkeley, California 94720, USA}
 \email{kdriver@berkeley.edu}
 \homepage{http://militzer.berkeley.edu/~driver/}

\author{F. Soubiran}
 \affiliation{Department of Earth and Planetary Science, University of California, Berkeley, California 94720, USA}

\author{Shuai Zhang}
 \affiliation{Department of Earth and Planetary Science, University of California, Berkeley, California 94720, USA}

\author{B. Militzer}
 \affiliation{Department of Earth and Planetary Science, University of California, Berkeley, California 94720, USA}
 \affiliation{Department of Astronomy, University of California, Berkeley, California 94720, USA}

\date{\today}

\maketitle
\vspace{-8in}
\section{Convergence Tests}

In this section, we provide raw data from our PIMC and DFT-MD
time-step and finite-size convergence
calculations. Table~\ref{EOSTableSuppTS} shows the results of static
path integral Monte Carlo (PIMC) calculations for a 8-atom as a
function of time-step for a fixed density. For a time-step of
0.00390625 Ha$^{-1}$, which we used in our production calculations,
the results are well converged. The pressure has 0.3\% error and
internal energy has 0.3\% error relative to the smallest time-step.

Table~\ref{EOSTableSuppFS} shows the comparison of pressures and
internal energies for a 24-atom and 8-atom simulation cell as a
function of temperature at a fixed density. Results are shown for both
PIMC and density functional theory molecular dynamics (DFT-MD).  The
absolute difference between the 8-atom and 24-atom pressures and
internal energies is only a fraction of a per cent of the total
values, and often within the statistical error. As expected in DFT,
the agreement between 8- and 24-atom results generally improves with
temperature. Above $1\times10^5$~K, the 8-atom cell size is sufficient
as the gamma-only k-point approximation becomes irrelevant.

\begin{table*}[!h]
\renewcommand\thetable{SI}
\caption{Convergence of oxygen energy and pressure with respect to
  PIMC time-step for static calculation of an 8-atom cell at a fixed
  density and temperature.}
\begin{tabular}{rrrrrr}
\hline
\hline
$\rho$ (g$\,$cm$^{-3}$)  & T(K)  &  Time-step (Ha$^{-1}$)   & P (GPa) & E (Ha/atom)\\
\hline
7.26176 & 1010479 & 0.015625    & 16700(45) & -51.30(4)\\
7.26176 & 1010479 & 0.0078125   & 17030(20) & -50.60(2)\\
7.26176 & 1010479 & 0.00390625  & 17200(30) & -50.17(3)\\
7.26176 & 1010479 & 0.00195312  & 17260(45) & -50.00(6)\\
\hline
\hline
\end{tabular}
\label{EOSTableSuppTS}
\vspace{-6pt}
\end{table*}

\begin{table*}[!h]
\renewcommand\thetable{SII}
\caption{Comparison of oxygen pressures and internal energies computed
  for 24- and 8-atom simulations cells as a function of temperature at
  a fixed density and their relative absolute (ABS) errors. The
  numbers in parentheses indicate the one-sigma statistical
  uncertainties of the DFT-MD and PIMC simulations.}

\begin{tabular}{rrrrrrrr}
\hline
\hline
$\rho$ (g$\,$cm$^{-3}$)    &  T (K)        & P (GPa)        & P (GPa) &  $\Delta$ P (GPa)   & E (Ha/atom)   & E(Ha/atom) & $\Delta$ E (Ha/atom)\\
            &            & 24-atom cell    & 8-atom cell    & ABS error   & 24-atom cell  & 8-atom cell &   ABS error\\
\hline
7.26176$^a$ & 1034730000 & 35139936(3207) & 35142831(2985) & 2895(4381) & 44226(4)   & 44227(4)   & 1(6)       \\
7.26176$^a$ & 16167700   & 537967(253)    & 538734(172)    & 767(305)   & 664.3(3)   & 664.3(2)   & 0.0(4)     \\
7.26176$^a$ & 8083850    & 262087(87)     & 261808(75)     & 279(115)   & 312.2(1)   & 311.4(1)   & 0.79(3)    \\
7.26176$^a$ & 2020960    & 49881(56)      & 49637(51)      & 245(78)    & 2.3(1)     & 1.74(7)    & 0.52(1)    \\
7.26176$^a$ & 998004     & 21158(46)      & 20964(31)      & 195(55)    & -44.95(5)  & -45.53(4)  & 0.59(7)    \\
7.26176$^b$ & 1000000    & 21387(48)      & 21301(20)      &  85(52)    & -46.11(6)  & -46.16(4)  & 0.05(7)    \\
7.26176$^a$ & 750000     & 15033(54)      & 15122(42)      &  88(69)    & -53.19(7)  & -53.17(5)  & 0.03(8)    \\
7.26176$^b$ & 750000     & 15272(29)      & 15236(21)      &  37(36)    & -54.49(3)  & -54.51(2)  & 0.01(3)    \\
7.26176$^b$ & 500000     & 9433(14)       & 9424(10)       &  9(17)     & -62.65(1)  & -62.652(7) & 0.00(1)    \\
7.26176$^b$ & 250000     & 4292(5)        & 4268(5)        &  24(7)     & -70.089(3) & -70.098(2) & 0.009(4)   \\
7.26176$^b$ & 100000     & 1831(3)        & 1765(6)        &  66(7)     & -73.613(2) & -73.663(4) & 0.050(4)   \\
7.26176$^b$ & 50000      & 1210(2)        & 1107(4)        &  103(4)    & -74.382(1) & -74.448(2) & 0.066(2)   \\

\hline
\hline
\end{tabular}
\label{EOSTableSuppFS}
\vspace{-6pt}
\begin{flushleft}
$^a$PIMC \\
$^b$DFT-MD\\
\end{flushleft}
\end{table*}